# Tribology of thin wetting films between a bubble and a moving solid surface

S.I. Karakashev[1], K.W. Stöckelhuber[2], R. Tsekov[1], C.M. Phan[3] and G. Heinrich[2]
[1]Department of Physical Chemistry, University of Sofia, 1164 Sofia, Bulgaria
[2]Leibniz Institute of Polymer Research, Dresden, D-01067 Dresden, Germany
[3]Curtin University, Department of Chemical Engineering, Perth, WA 6845, Australia

The tribology of a bubble rubbing on a solid surface is studied via interferometry. A unique experimental setup is designed for monitoring the thickness profiles of a wetting film, intercalated between the bubble and hydrophilic glass moving with speed up to 412 μm/s. The determination of the 3D film thickness profiles allows us to calculate 3D maps over the wetted surface of the local capillary, disjoining and lift pressures, viscous stress and friction force. In this way the average friction force and the corresponding friction coefficient are obtained. A theoretical model for the dependence of the friction coefficient on the film thickness is developed. The relevant slip coefficient, being a measure for the slip between liquid and solid, is determined as a function of the speed of the solid surface. It is found out that below 170 μm/s a friction regime exists which formally resembles dry friction, while at larger speed the friction force between the bubble and solid passes through a maximum. Furthermore, the friction coefficient has a large value at low speed of the solid and reduces substantially with the speed increase.

The science of friction traces its roots five centuries back to the pioneering works of da Vinci[1]. Amontons[2], Euler[3] and Coulomb[4] had later established the fundamental laws of dry friction between sliding solid bodies. Further on the dry contact between elastic bodies has been investigated by Hertz[5], who laid out the foundations of contact mechanics. Reynolds[6] found out that a liquid layer, intercalated between two solid surfaces in relative motion towards each other, becomes a lubricant. Thus, he has developed the hydrodynamic lubrication theory, which found many useful applications.[7-11] With the advancement of technology, the friction is studied nowadays on micro- and nano-levels[12,13] indicating elastic and plastic deformations of the surface asperities during the sliding process. The seminal work of Hertz[5] on the friction between sliding elastic surfaces has been further developed in the middle of the previous century,[14,15] while the contribution of adhesion between the elastic bodies in contact has been accounted for in the Johnson-Kendall-Roberts[16] (JKR) and the Derjaguin-Muller-Toporov[17] (DMP) theories. The term tribology is suggested by Jost[18] as "the science and technology of interacting surfaces in relative motion and practices related to". The term nano-tribology appeared upon bringing the scale of the investigations of sliding surfaces to nano and atomic dimensions.[8,19,20] At present knowledge about nano-tribology is exploited for designing micro- and nano-electromechanical systems (MEMS/NEMS), self-lubricating and biologically inspired surfaces.[21]

The tribological studies mentioned above are focused on sliding solid phases in dry or lubricated contact. Thin wetting film is a nano-layer lubricating the friction between a bubble and a solid surface when they are in relative motion towards each other. In addition, the thin layer is squeezed, due to the load force pressing the bubble toward the solid surface. The resulting liquid outflow is laminar and its hydrodynamics obeys the lubrication approximation.[6,22-26] The theoretical problem is even more complicated as far as the bubble surface is deformable and an additional slip on the solid surface could also appear. Thin films between rubber ball and moving solid have been already studied. It is reported[27] that the load force squeezes out the liquid layer between the two surfaces until the point of touch. Upon moving the solid surface a lift force emerges tending to separate the two surfaces. The lift force increases with enhancing the speed of the solid until the point, where the load and lift forces become equal to each other and the rubber ball starts hydroplaning. The detaching of the two surfaces of the film occurs at a certain speed of motion of the solid. This valuable paper is a part of series of works[27-30] devoted to the mechanism of rupturing of the film intercalated between elastomeric ball and moving solid. Except the mechanism of rupturing and the critical velocity of hydroplaning, which are essential for such phenomena, the dependence of the friction and lift forces on the speed of motion of solid is important for the proper understanding of the hydroplaning of elastic surfaces. Being inspired by these works, we build up a unique experimental setup for investigating the tribology of bubble rubbing on solid surface. We replaced the elastomeric ball with a bubble in our studies. The utilization of such kind of experimental setup is valuable from fundamental viewpoint because the bubble is substantially more deformable than the elastomeric ball, which makes us able to determine the 3D film thickness profile and subsequently 3D maps of the lift and friction forces over the surface of the bubble. Moreover, thin wetting films on moving solid surface have been never investigated despite the intensive research on them since their very early studies in 1939.[31,32] In addition to the fundamental viewpoint such study will be of interest for specialists in flotation of mineral particles, whose object of research is the interaction between particle and bubble under dynamic conditions.

The experimental setup specially designed for the experiments consist of the following elements (Fig. 1): i) table allowing automated motion in *x* and *y* directions with different speed; ii) experimental cell filled with aqueous medium; iii) capillary tube filled with air and connected hermetically with micro-syringe for increasing the tube pressure; iv) pressure transducer; v) objective of upright microscope working under regime of reflection. The capillary tube is immersed into the experimental cell. The pressure inside the cell is increased by means of the micro-syringe. This forms a bubble on the top of the capillary tube. The bubble gets in contact with the substrate, situated on the bottom of the experimental cell. Thus a thin wetting film is formed and the load pressure between the bubble and the substrate surface is measured by means of the pressure transducer. The two surfaces of the wetting film are illuminated normally with polychromatic and coherent light through the microscope objective. The reflected light by the two film surfaces forms an interference pattern, which is stored to computer and processed offline. The lower surface of the wetting film (the solid substrate) is moved tangentially by means of the automatic moving table. Thus the experimental cell with the solid substrate on its bottom is moved in *x* or *y* direction, while the capillary tube with the bubble on its top remains stationary.

The offline processing of the interference pattern provides information about the variation of the film thickness profile in a given time interval. Thus the load pressure, the local thickness in every point of the wetting film and the tangential speed of the lower surface are monitored. The experiment are conducted with speed of the solid surface in wide range of 6 to 412 μm/s. At each speed of motion of the solid substrate, the wetting films have particular stationary shape corresponding to a certain 3D film thickness profile. A movie for each particular case is made followed by off-line processing by means of Image J program. To achieve a 3D profile, a stationary interferogram of a certain frame of the movie was chosen.

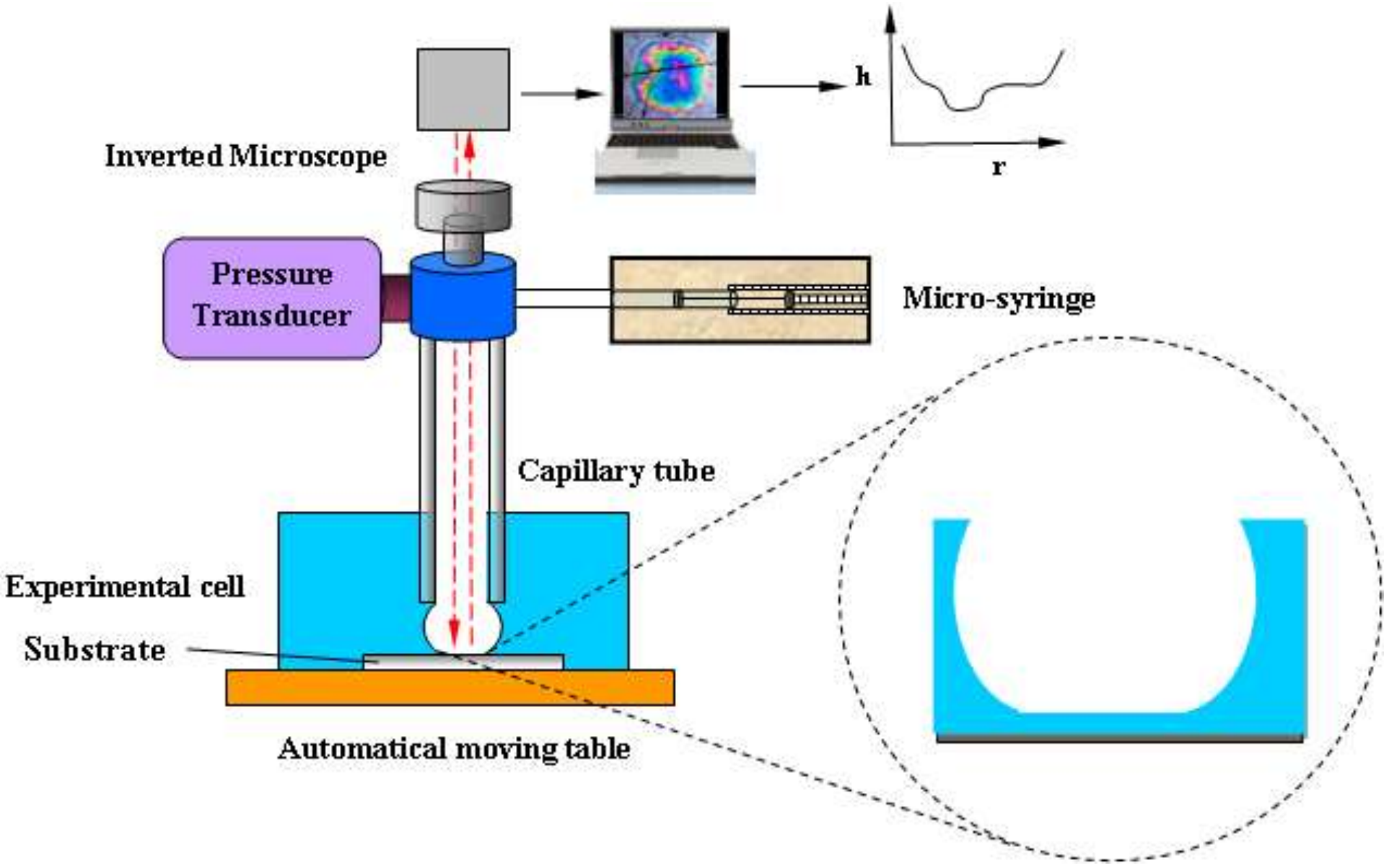

**Figure 1.** Experimental setup for tribological experiments on wetting films.

Each picture of the interferogram has been processed in the following way (see Fig. 2): Multiple horizontal and parallel to each other lines crossing the film have been drawn by means of the graphical tool of Image J software. Each of these lines, coinciding with the horizontal *x* axis, was coupled with particular spatial interferogram along the line crossing the film, obtained by “plot profile” option of Image J software. Thus, a series of spatial interferograms corresponding to each line along the axis *x* were obtained and converted to real film thickness profiles by means of the interferometry formula[24]. The distances, coinciding with the vertical axis *y* between the lines, were known as well. Thus were obtained the exact coordinates *x* and *y* corresponding to each film thickness. The 3D profiles were obtained by 100 x 100 matrixes via Microcal Origin software version 5. We have exploited the 3 D film thickness profile to obtain 3D maps of the local capillary pressure, disjoining pressure, viscous dissipation, lift pressure, and the local friction force over the wetting film surface. Thus, we obtained the averaged values of these parameters as a function of the speed of motion of the solid substrate.

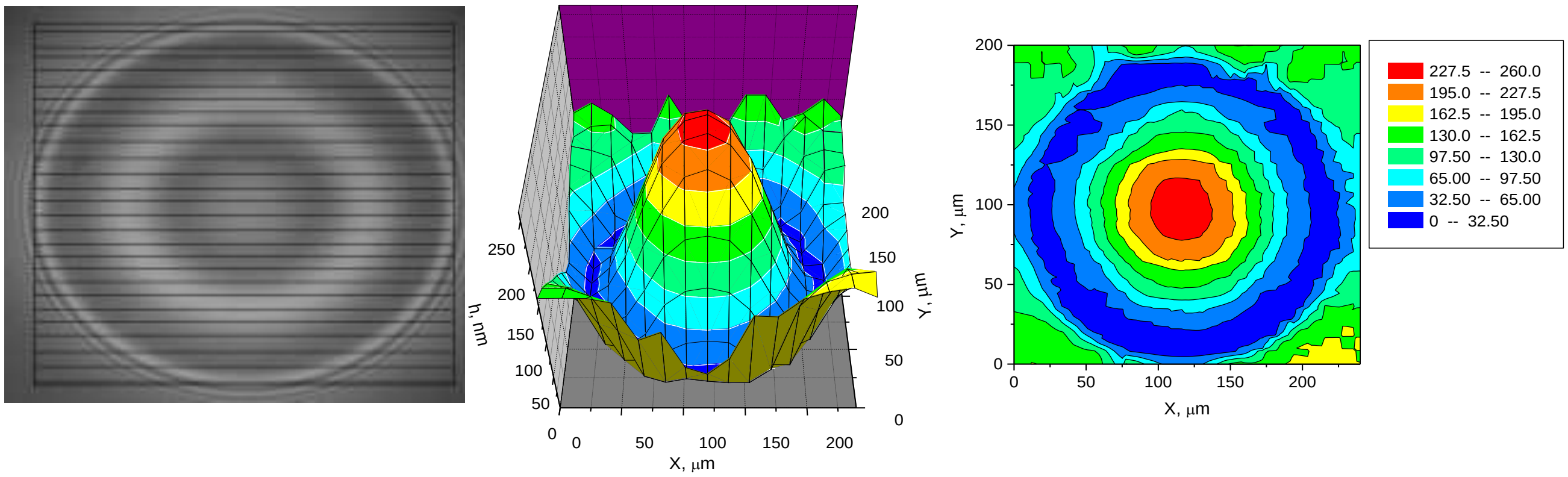

**Figure 2.** Processed interferogram of dimple-like wetting film and the corresponding 3D film thickness profile.

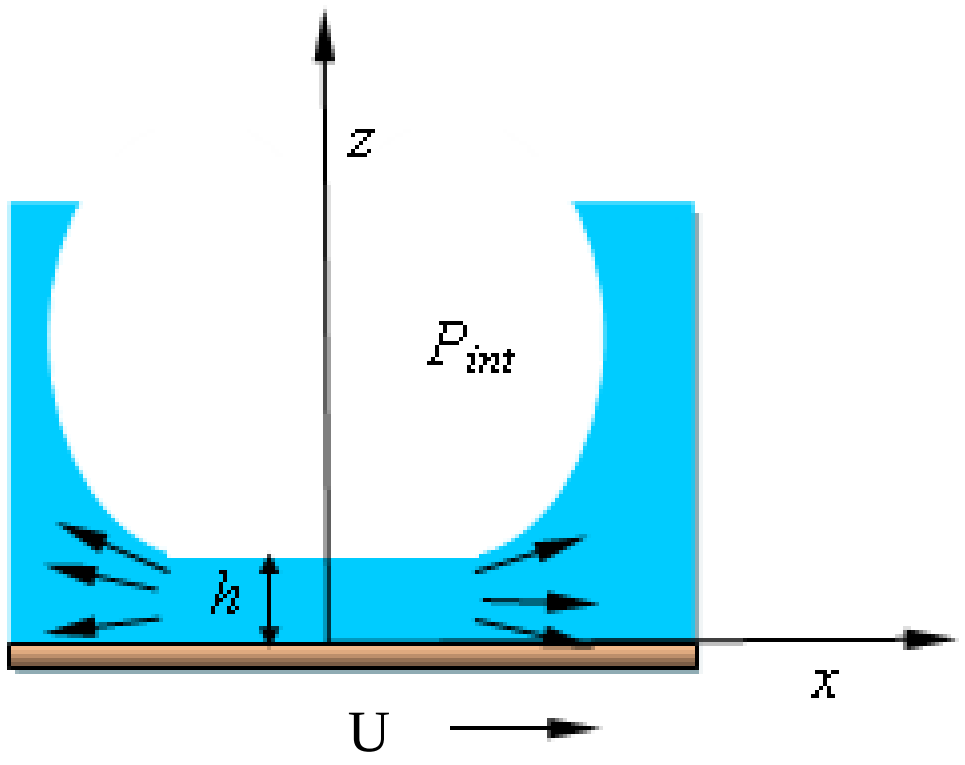

**Figure 3.** Sketch of a wetting film pressed by the load gas pressure to a solid surface moving with speed $U$

The physics behind the bubble rubbing on solid surface is simple. There is a constant capillary pressure pushing the bubble toward the solid surface and the latter is opposed by the disjoining pressure. When these two pressures are equal, the wetting film is planar at equilibrium. If the solid surface starts moving, a pressure gradient along the film emerges making the film inhomogeneous and thicker, which results in a weaker and inhomogeneous disjoining pressure. The bending of the film air/water interface generates local capillary pressure tending to recover the original planar shape. In addition, there is a viscous dissipation across the film. When all these effects are balanced the corresponding wetting film is stationary. Let us have a wetting film with thickness *h* and the origin of the Cartesian coordinate system is shown in Fig. 3. The lower solid surface of the film moves in the *x* direction with controlled speed *U*. In the frames of the Reynolds lubrication approximation the Navier-Stokes equations can be presented in the following form

$$\partial_x p = \mu \partial_z^2 v_x \qquad \partial_y p = \mu \partial_z^2 v_y \qquad \partial_z p = 0 \qquad (1)$$

Here $p$ is the local film pressure, $\mu$ is the dynamic viscosity, $v_x$, $v_y$ and $v_z$ are the three components of the hydrodynamic velocity in the film, respectively. The relevant boundary conditions to these equations are

$$\mu(\partial_z v_x)_{z=0} = \beta(v_x - U)_{z=0} \qquad \mu(\partial_z v_y)_{z=0} = \beta(v_y)_{z=0} \qquad (\partial_z v_x)_{z=h} = (\partial_z v_y) = 0 \qquad (2)$$

The last equation corresponds to the stress-free air/water interface, while the first two Navier equations[35] presume a possible slip on the glass/water surface with a slip friction coefficient $\beta$. The usual nonslip boundary condition corresponds to $\beta = \infty$, while $\beta = 0$ is valid for stress-free surfaces as the air/water one. The ratio $\mu / \beta$ is known as the slip length. From a previous investigation of an oscillating bubble against a quartz plate[37] one can determine a value of the relevant slip coefficient of $\beta = 1.25$ kPa s/m, which corresponds to a slip length of 800 nm. It is important to stress that the slip length is not directly related to the slipping plane, which is attribute of an integral approximation of the overall hydrodynamics. In the case of presence of slip effect, the zeta potential will be probably affected by the value of the slip coefficient $\beta$. Note that the latter is very sensitive to entrapped air on the glass/water surface (the Cassie effect) and for this reason the super-hydrophobic surfaces possesses huge slip lengths larger than 10 µm.[38] Integrating Eq. (1) under the boundary conditions (2) yields

$$v_x = U - [(2h - z)z / 2\mu + h / \beta]\partial_x p \qquad v_y = -[(2h - z)z / 2\mu + h / \beta]\partial_y p \qquad (3)$$

Introducing these velocity components in the continuity equation $\partial_x v_x + \partial_y v_y + \partial_z v_z = 0$ results, after integration under the constrain $(v_z)_{z=0} = 0$, to the normal component of the hydrodynamic velocity

$$v_z = z\partial_x\{[(3h - z)z / 6\mu + h / \beta]\partial_x p\} + z\partial_y\{[(3h - z)z / 6\mu + h / \beta]\partial_y p\} \qquad (4)$$

Using the kinematic boundary condition $\partial_t h + v_x \partial_x h + v_y \partial_y h = v_z$ at $z = h$ one can obtain from Eqs. (3) and (4) the following dynamic equation

$$\partial_t h = \partial_x[(h / 3\mu + 1 / \beta)h^2 \partial_x p - Uh] + \partial_y[(h / 3\mu + 1 / \beta)h^2 \partial_y p] \qquad (5)$$

To describe completely the evolution of the wetting film thickness Eq. (5) requires an expression for the film pressure, which is given by the normal force balance on the air/water interface

$$p = p_\sigma - \sigma(\partial_x^2 h + \partial_y^2 h) - \Pi(h) \tag{6}$$

Here $p_\sigma$ is the constant capillary pressure of the bubble, $\sigma$ is the surface tension on the air/water interface of the film and the disjoining pressure $\Pi(h)$ can be expressed by the DLVO theory

$$\Pi = \Pi_{EL} + \Pi_{VW} = 64ck_B T \tanh(e\varphi_1 / 4k_B T)\tanh(e\varphi_2 / 4k_B T)\exp(-\kappa h) - A / 6\pi h^3 \tag{7}$$

$\Pi_{EL}$ and $\Pi_{VW}$ are electrostatic and van der Waals components, $c$ is the electrolyte concentration, $T$ is temperature, $\varphi_1 = -65\ \mathrm{mV}$ and $\varphi_2 = -42\ \mathrm{mV}$ are the relevant surface potentials[33] on the air/water and glass/water interfaces, respectively, $A = -1\times 10^{-20}\ \mathrm{J}$ is the Hamaker constant[34] and $\kappa = (2e^2 c / \varepsilon_0 \varepsilon k_B T)^{1/2}$ is Debye parameter with $\varepsilon_0\varepsilon$ being the dielectric permittivities of water. The combination of Eqs. (5), (6) and (7) is a nonlinear system, which can be solved only numerically if the initial profile is known.

To get inside in the problem let us consider now the case of a stationary thickness distribution with $\partial_t h = 0$. If in addition the film is strictly symmetric, the film pressure will depend on $x$ only and Eq. (5) will simplify after integration to

$$h\partial_x p = 3\mu U / (h + 3\mu / \beta) \tag{8}$$

The friction force acting on the solid surface can be calculated via the expression

$$f = \oint_{\pi R^2} \beta(U - v_x)_{z=0}\, dxdy = \oint_{\pi R^2} h\partial_x p\, dxdy \tag{9}$$

where the last result follows from Eq. (3). Substituting in the simplified case the pressure gradient in Eq. (9) from Eq. (8) yields that the friction force is proportional to the velocity $f = bU$ and the friction coefficient is given by the following equation

$$b = \oint_{\pi R^2} \frac{3\mu}{h + 3\mu / \beta}\, dxdy \tag{10}$$

where $R$ is the film radius. In the case on homogeneous film with nonslip surface ($\beta = \infty$) Eq. (10) reduces to the Lorentz formula $b = 3\pi\mu R^2 / h$, which provides the Stokes law $b = 6\pi\mu R$ when $h = R/2$ In the opposite case of small $\beta$ Eq. (10) tends to $b = \beta\pi R^2$, which is independent of the film thickness. Note that this expression, which is independent of the film viscosity also, is always valid in the limit of very thin films. Due to the fact that the friction coefficient $b$ depends on the film radius, which can change at different measurements, it is better to introduce the friction coefficient surface density via the expression $\gamma \equiv b / \pi R^2 \approx (\bar{h} / 3\mu + 1/\beta)^{-1}$, where $\bar{h}$ is the average film thickness. Equation (9) can be used for building of a friction force map over the film surface. From an experimentally measured 3D film thickness profile, one can calculate the pressure field by Eqs. (6) and (7) and the friction force field via Eq. (9). Than one can calculate the average friction force per unit area

$$p_f = \frac{f}{\pi R^2} = -\frac{1}{\pi R^2} \oint_{\pi R^2} h \frac{\partial}{\partial x} [\sigma(\partial_x^2 h + \partial_y^2 h) + \Pi(h)] dxdy = \gamma U \qquad (11)$$

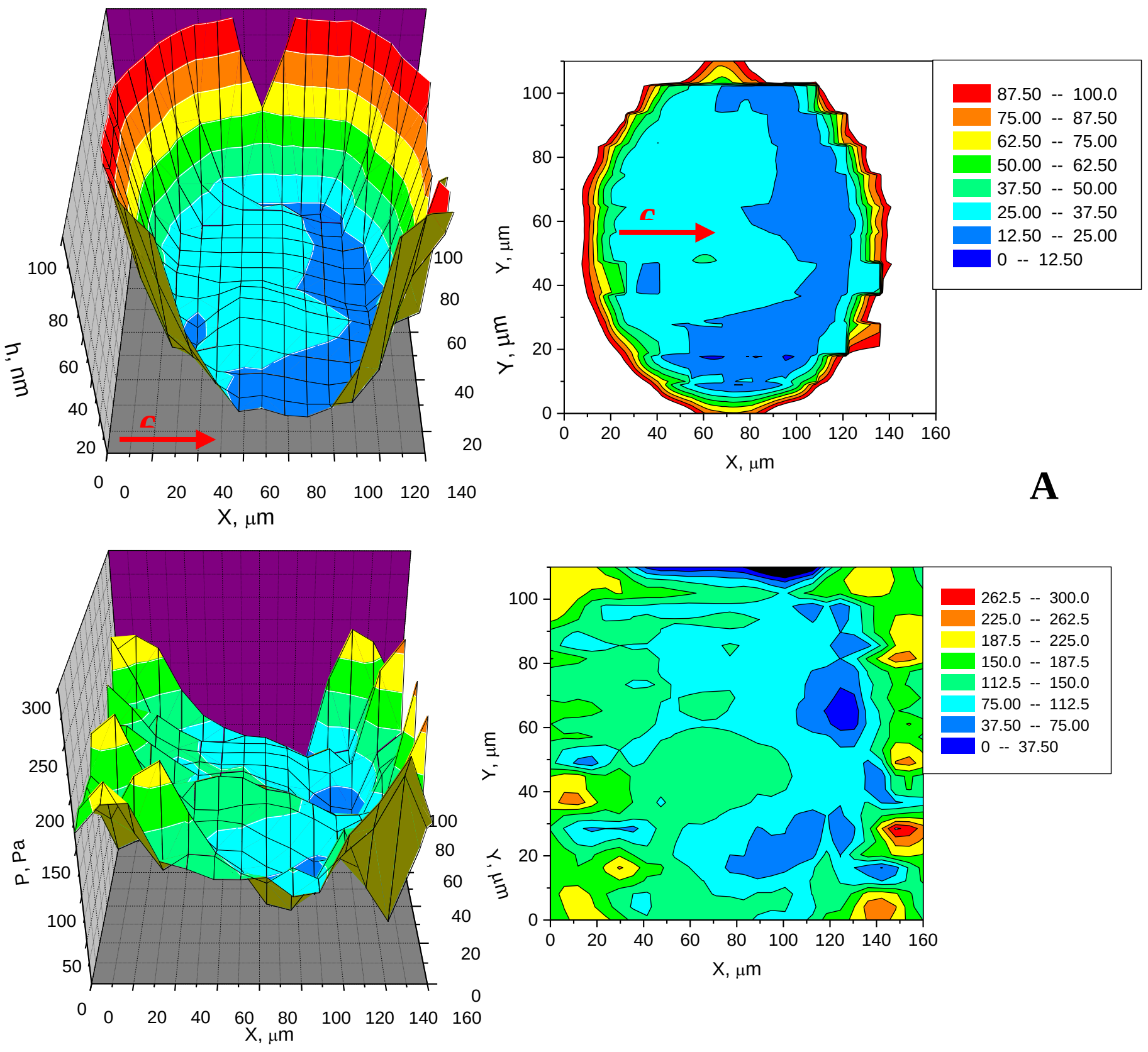

**Figure 4.** An example of 3D film thickness profile (upper graphs) and their corresponding lift pressure maps (lower graphs) for: A – 6 μm/s, B – 68 μm/s and C – 111 μm/s.

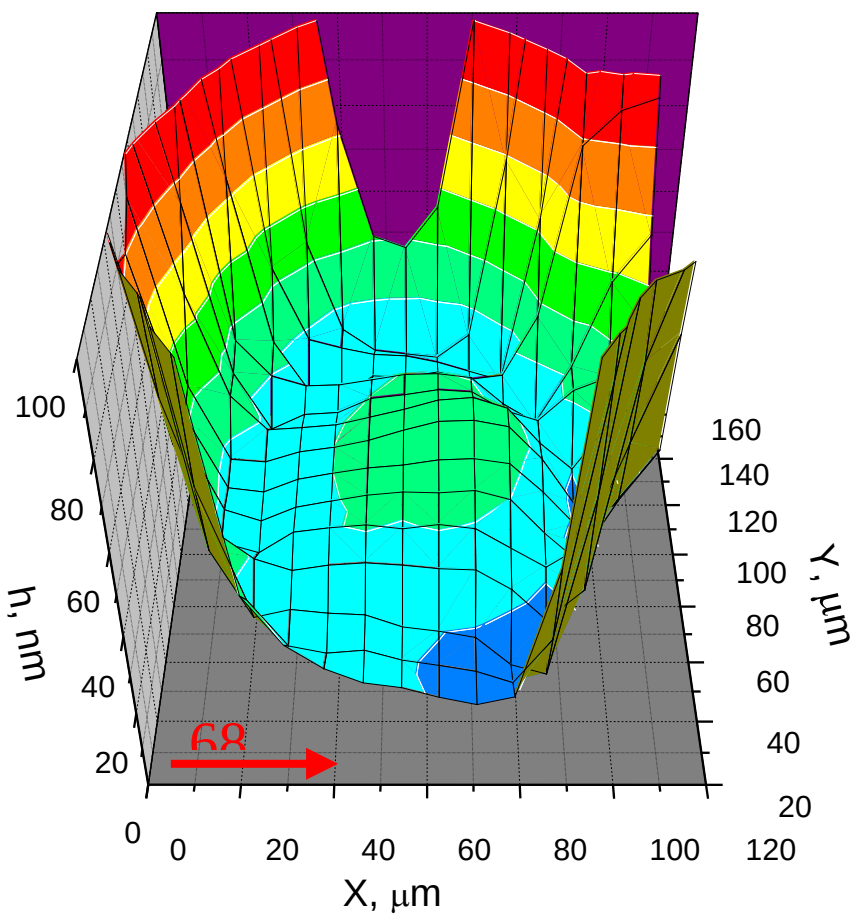
h, nm
X, µm
Y, µm
68

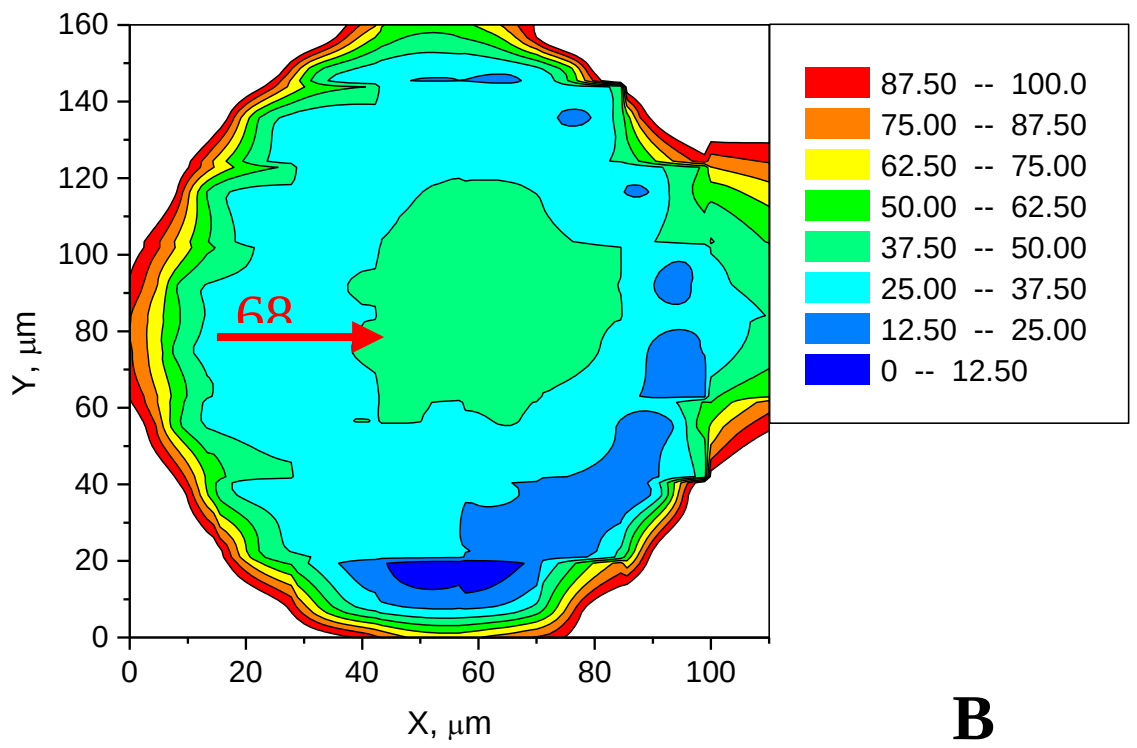
87.50 -- 100.0
75.00 -- 87.50
62.50 -- 75.00
50.00 -- 62.50
37.50 -- 50.00
25.00 -- 37.50
12.50 -- 25.00
0 -- 12.50
X, µm
Y, µm
68

**B**

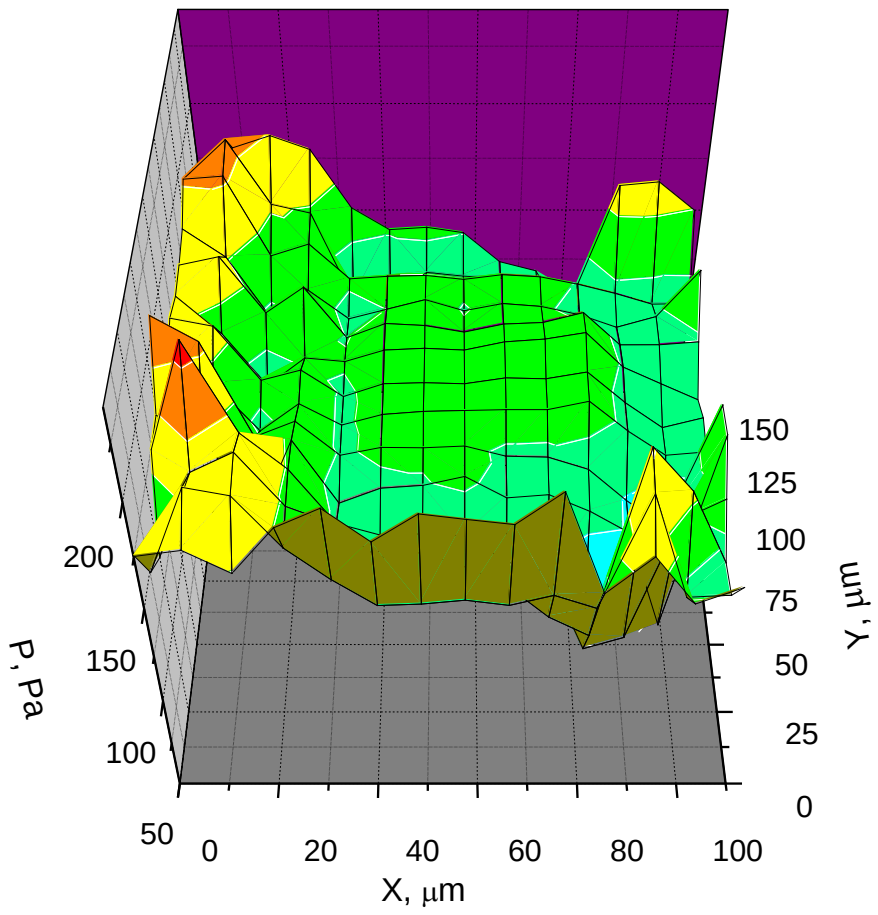
P, Pa
X, µm
Y, µm

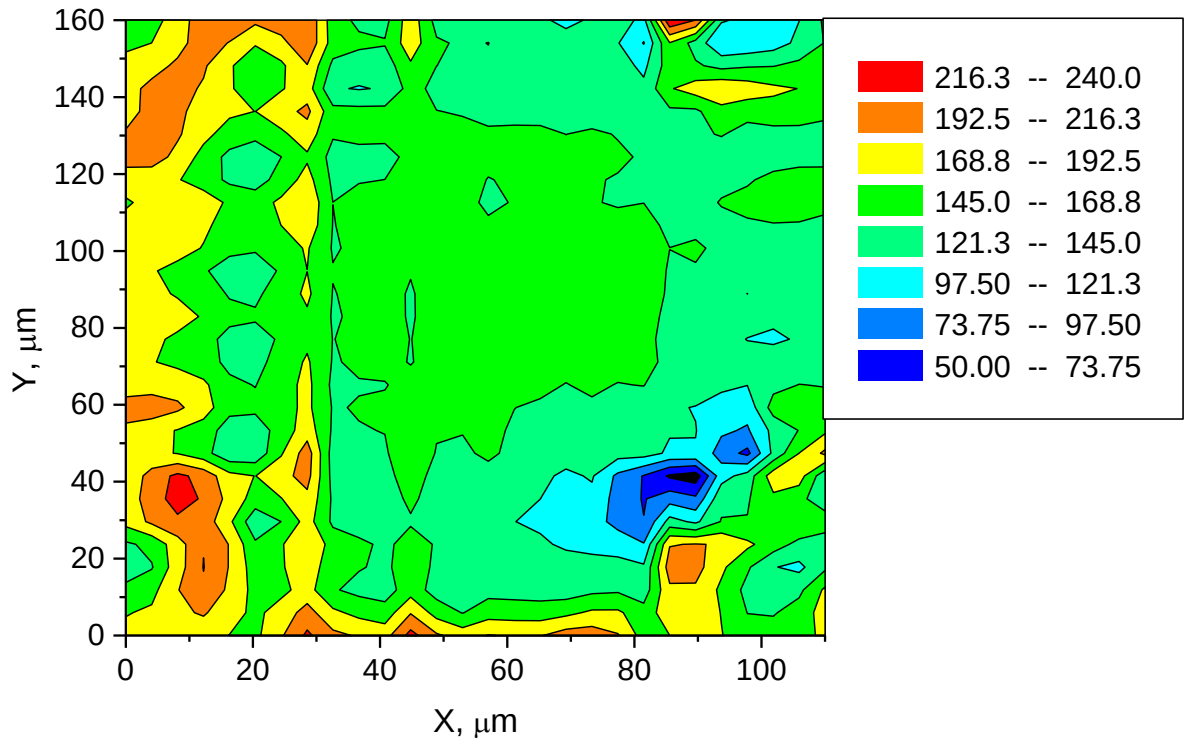
216.3 -- 240.0
192.5 -- 216.3
168.8 -- 192.5
145.0 -- 168.8
121.3 -- 145.0
97.50 -- 121.3
73.75 -- 97.50
50.00 -- 73.75
X, µm
Y, µm

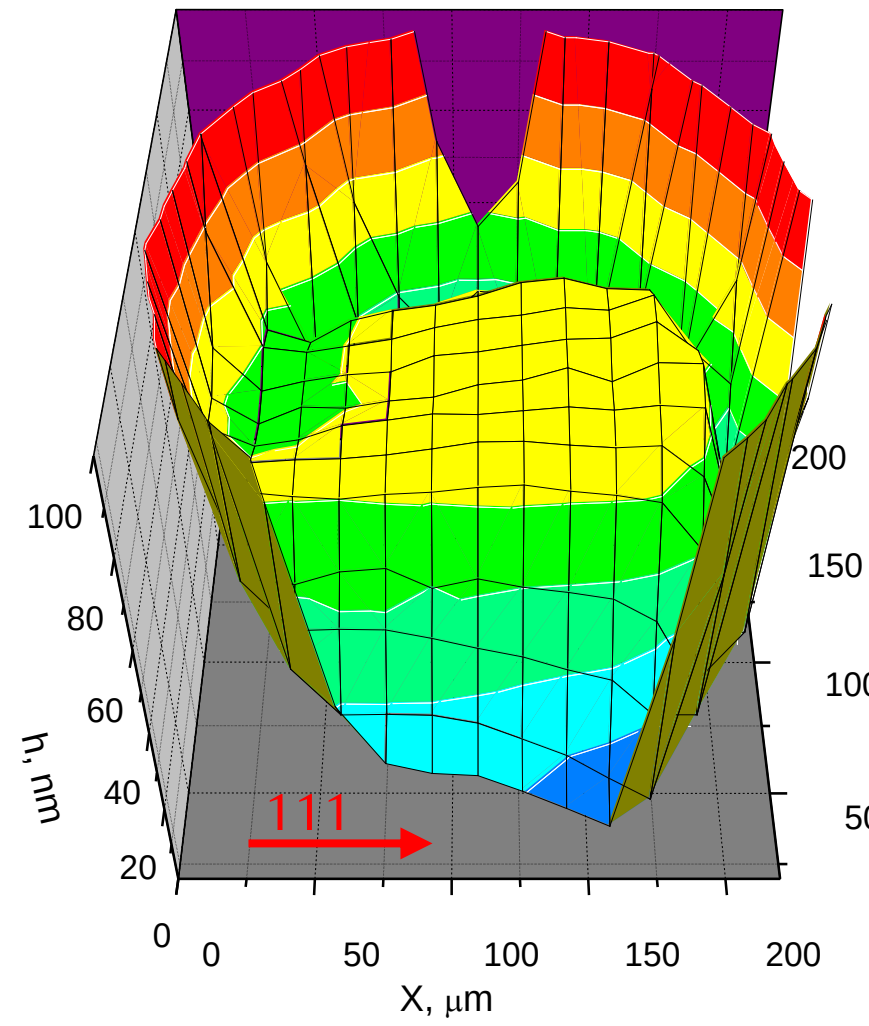
h, nm
X, µm
Y, µm
111

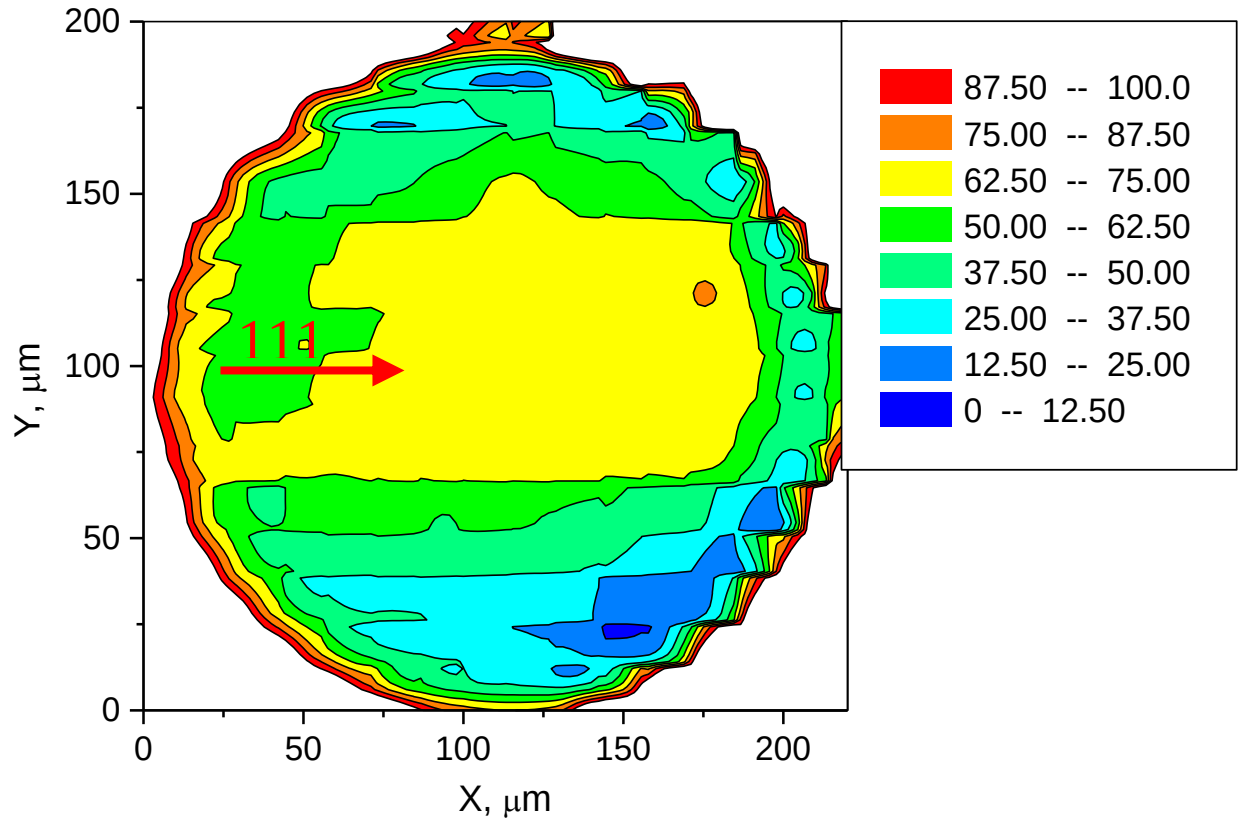
87.50 -- 100.0
75.00 -- 87.50
62.50 -- 75.00
50.00 -- 62.50
37.50 -- 50.00
25.00 -- 37.50
12.50 -- 25.00
0 -- 12.50
X, µm
Y, µm
111

**C**

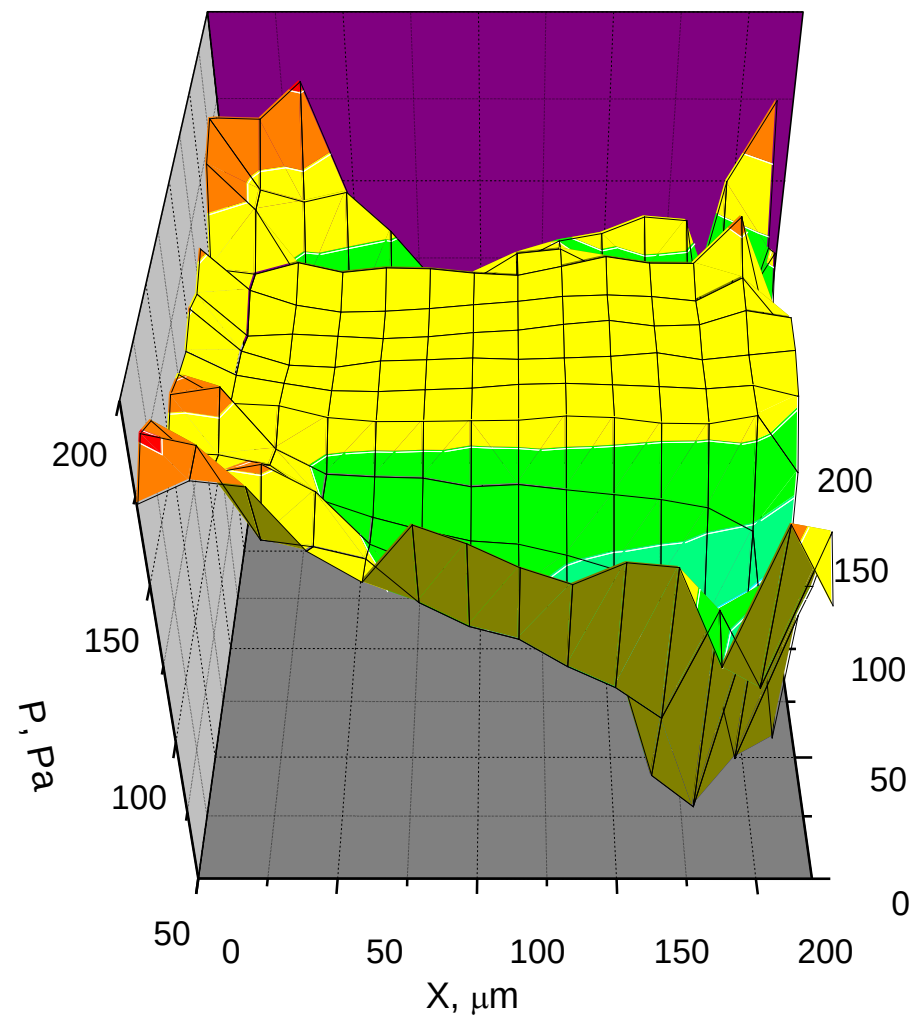
P, Pa
X, µm
Y, µm

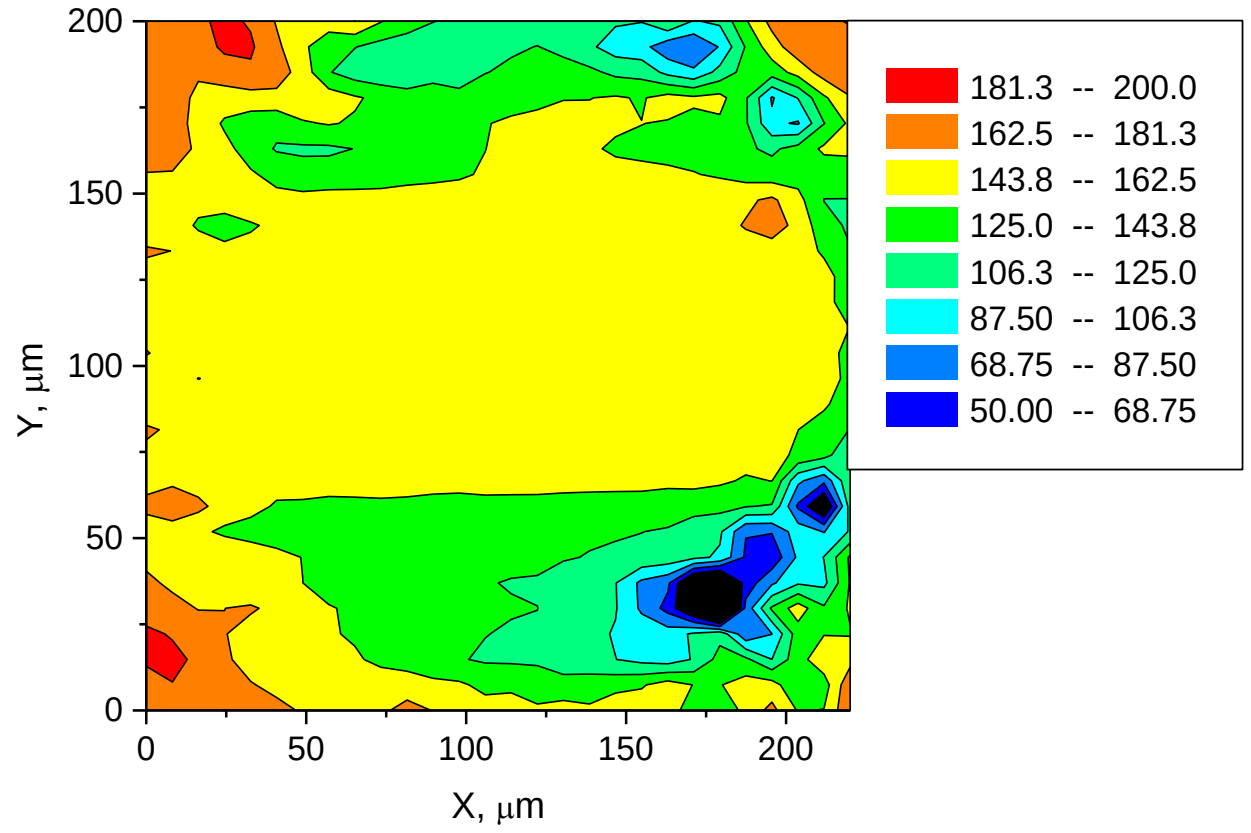
181.3 -- 200.0
162.5 -- 181.3
143.8 -- 162.5
125.0 -- 143.8
106.3 -- 125.0
87.50 -- 106.3
68.75 -- 87.50
50.00 -- 68.75
X, µm
Y, µm

We have processed the interferograms of stationary wetting films in the speed range of 6 to 412 μm/s and determined their 3D film thickness profiles (see Fig. 4). By means of these data we have determined and 3D maps of the local capillary pressure, disjoining pressure, viscous dissipation, hydrodynamic lift pressure and the local friction force over the wetting film surface. Figure 4 presents just an illustration of 3D film thickness profiles and their corresponding 3D maps of the lift pressure calculated by means of Eq. (11).

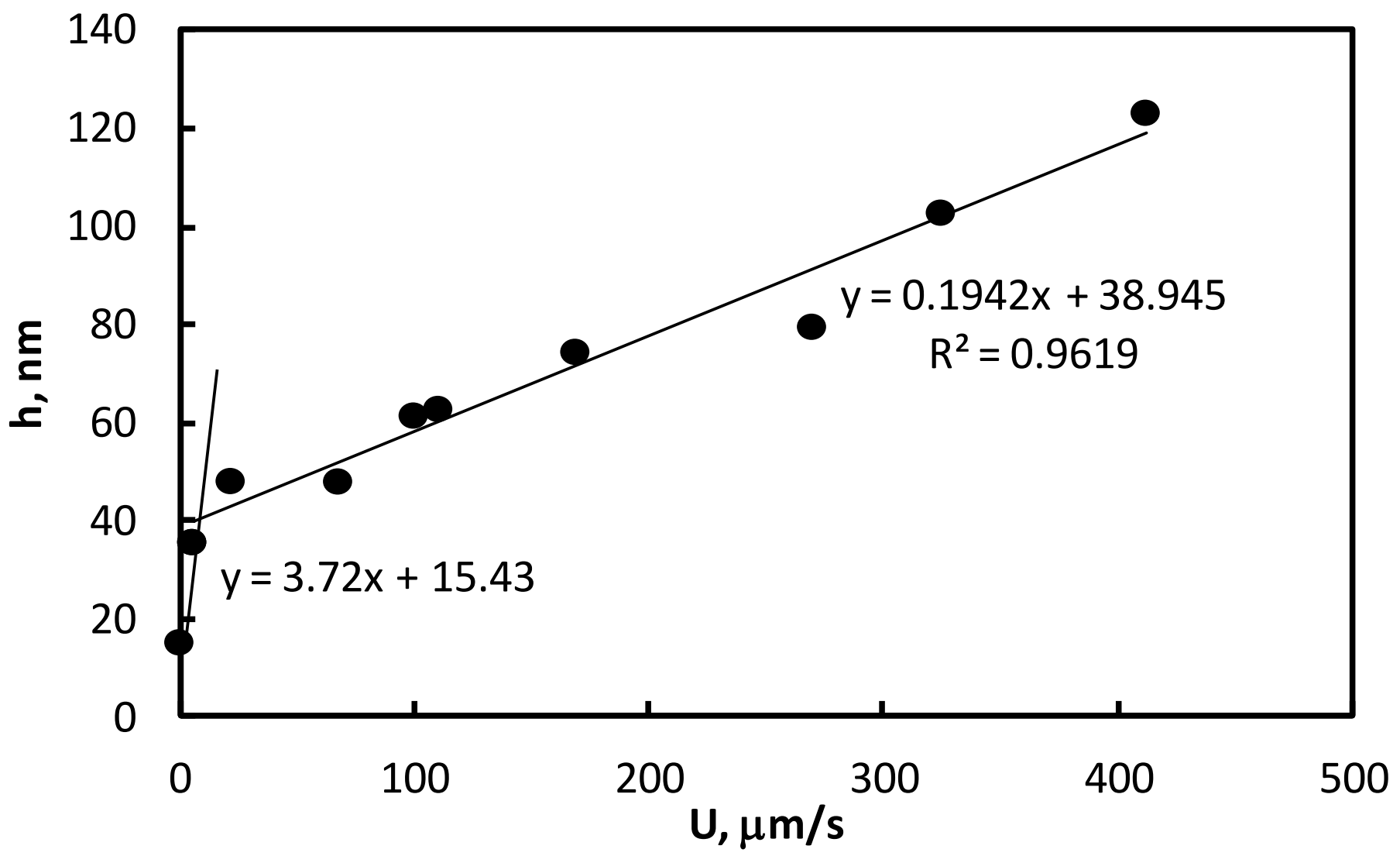

**Figure 5.** The average film thickness versus speed of the solid.

Figure 5 presents the average film thickness versus the speed of motion of the solid substrate. In the case of immovable solid substrate the wetting films are almost planar with equilibrium film thickness about 15.4 nm, which is due to the capillary pressure ( $p_\sigma \approx 240\ \mathrm{Pa}$ ) pushing the bubble towards the solid and the total disjoining pressure acting in opposite direction. One can see very steep increase in the average film thickness upon enhancing the speed of the solid motion from zero to about 15 μm/s, beyond which this dependence weakens about 19 times (see Fig. 5). The thickening of the wetting films is due to the additional lift pressure generated by the motion of the solid thus making the films hydroplaning. The dependence of the lift pressure on the speed of motion of the solid is presented in Fig. 6. Similarly to Fig. 5, the lift pressure increases steeply upon enhancing the speed of the motion from zero to about 15 nm, beyond which the dependence weakens about 144 times (see Fig. 6). The dependence of the average friction force per unit area on the speed of motion of the solid is presented in Fig. 7. One can see that within the range of speeds (0 – 412μm/s) this dependence passes through a maximum located at about 270 μm/s. Moreover, the friction between the bubble and the solid surface in range of 6 to 110 μm/s practically does not change (see Fig.7), which coincides with the third law of Amontons for the dry friction stating that the kinetic friction is independent of the sliding velocity. Yet, this friction depends on the apparent contact area between the bubble and solid (the film radius), which is in violation of the second law of Amontons stating that the force of friction is independent of the apparent area of contact. This promotes the idea that in the range of 6 to 110 μm/s a mixed regime between dry and lubricated friction.

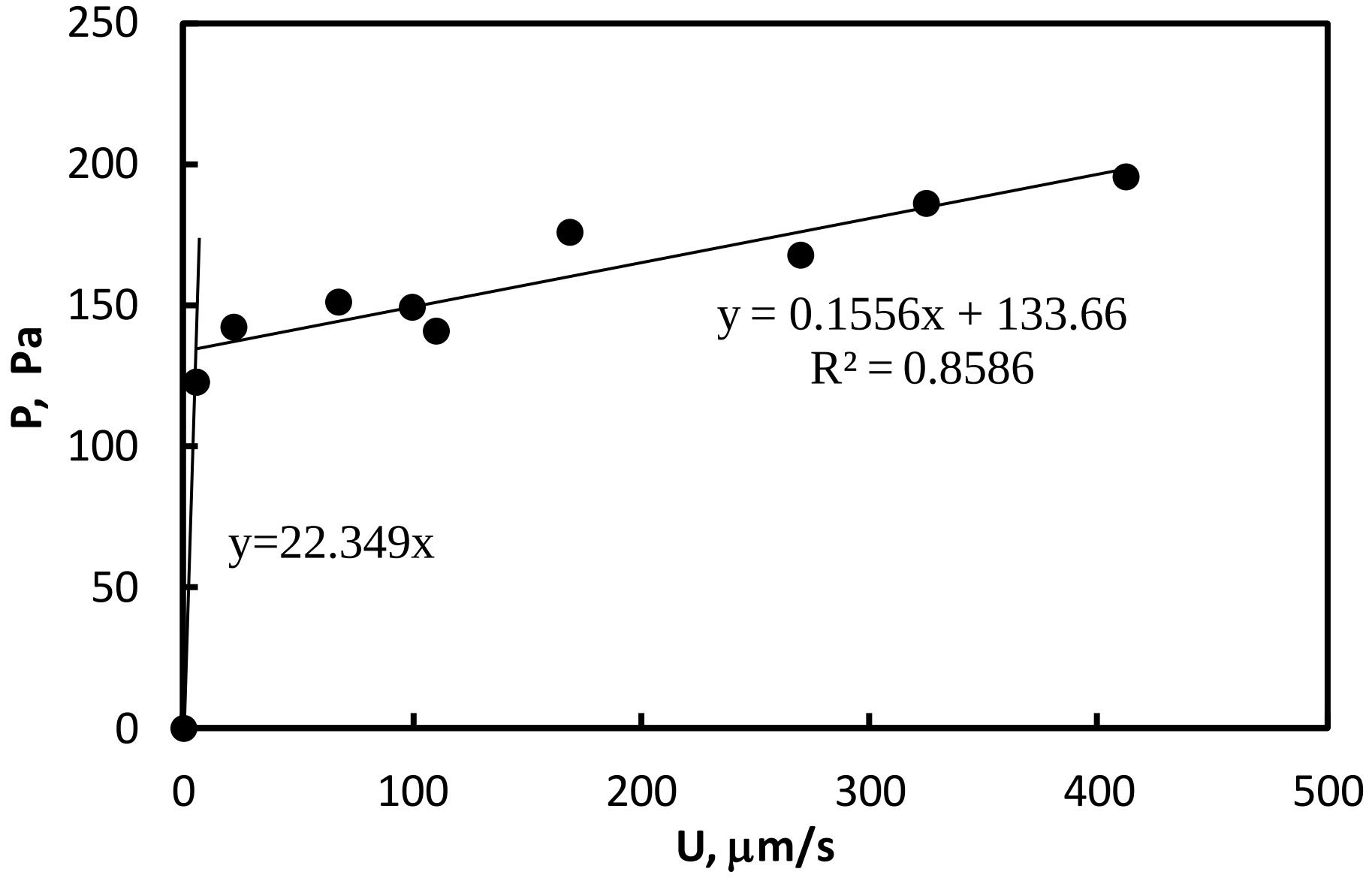

**Figure 6.** The averaged lift pressure versus speed of the solid.

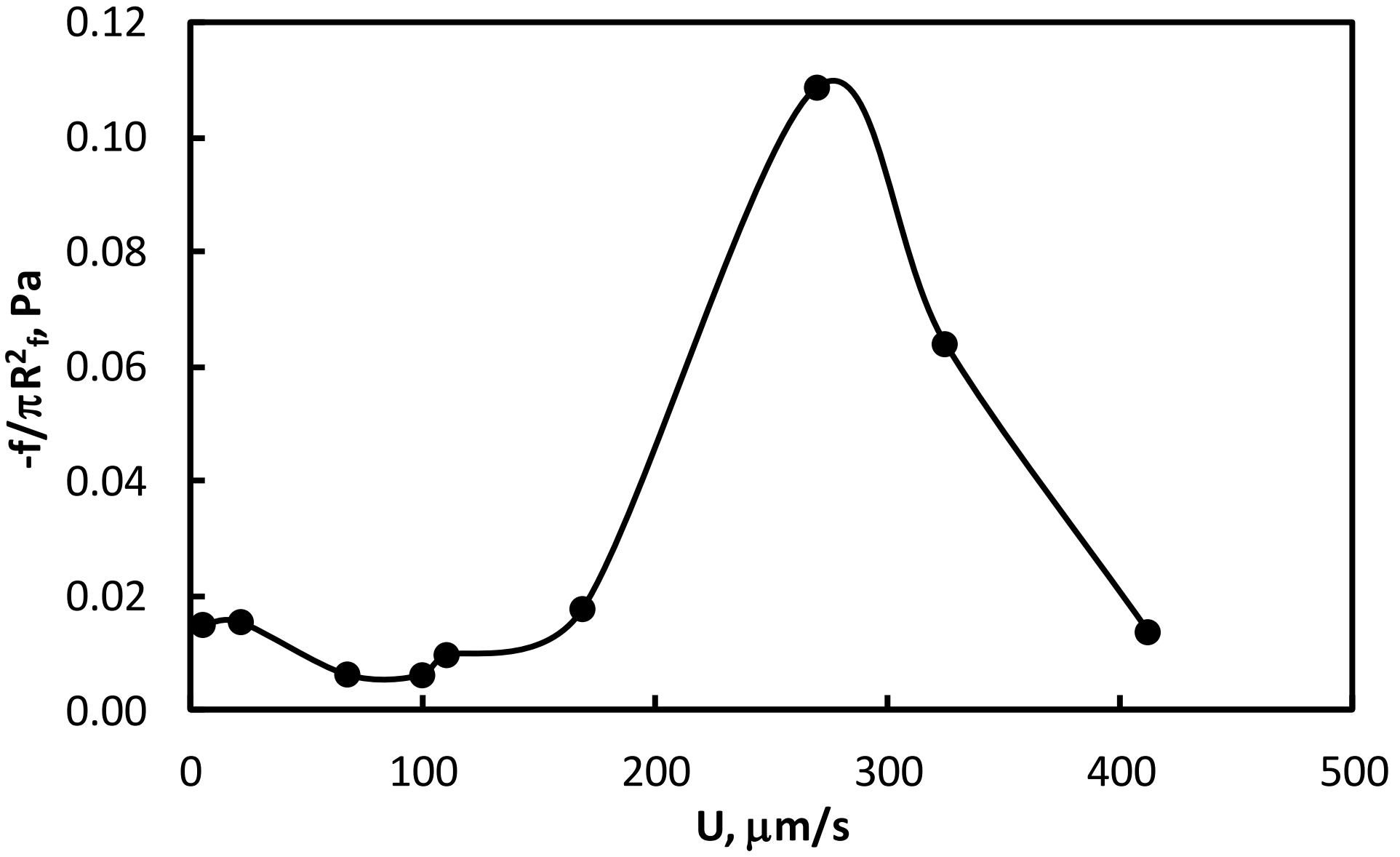

**Figure 7.** The averaged friction force per unit area calculated by means of Eq. (11) versus the speed of the solid.

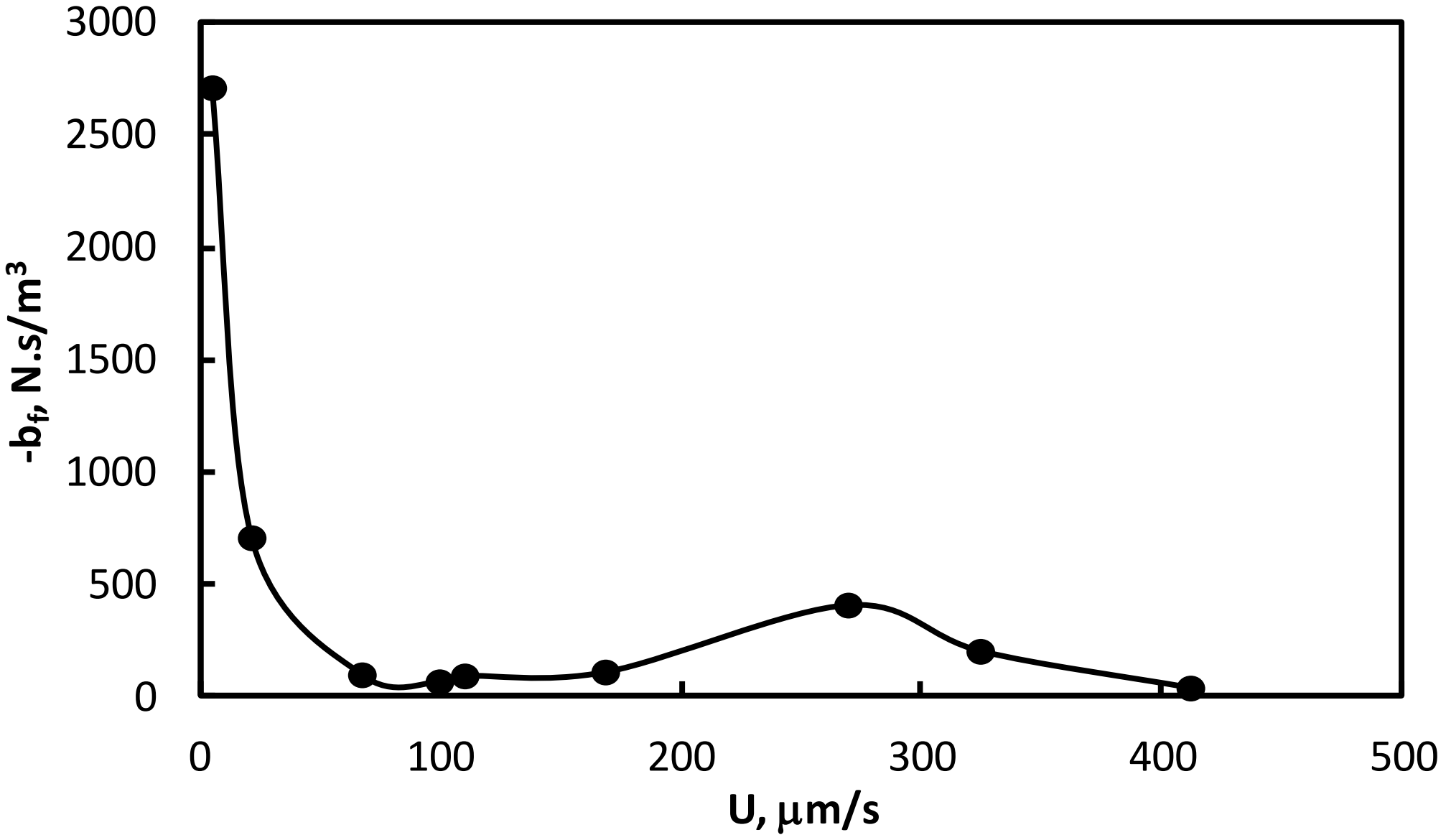

**Figure 8.** The specific friction coefficient per unit area calculated by means of Eq. (11) versus the speed of the solid.

The dependence of the friction coefficient per unit area $\gamma$ on the speed of solid motion is presented in Fig. 8. As seen in the figure, the values of the friction coefficient $\gamma$ substantially decreases upon increasing the speed of solid's motion up to 110 μm/s, beyond which does not change significantly with the further enhance of the speed of solid's motion Meanwhile, Eq. (10) is one of the central equations of our theoretical model because it describes how the friction coefficient $\gamma$ depends on the film thickness. An important parameter in Eq. (10) is the friction slip coefficient $\beta$, which has been defined in boundary conditions Eq. (2). If we suppose $\beta$ as a constant Eq. (10) predicts theoretical curve $\gamma$ versus $h$, which differs significantly from the experimental data. For this reason we concluded that $\beta$ is implicit function of the speed of solid motion *U*. Due to the fact that we do not have any theoretical description about $\beta$ as a function of *U*, we suggest the following empirical relation between the these two parameters:

$$\beta = \alpha / U^n \qquad (12)$$

where $\alpha$ and $n$ are unknown constants. We have applied Eq. (12) on the experimental points shown in Fig. 9 using the least square method and the Solver option of Microsoft Excel to obtain the two free parameters $\alpha$ and $n$. We obtained in this way the following empirical equation:

$$\beta = 1/65U \qquad (13)$$

where $U$ is given in μm/s. Figure 9 presents $\gamma$ versus $h$ experimental points (see Eqs. (9) and (11)) and the theoretical curve obtained by means of Eq. (10) and the empirical relation β=1/65*U*. The first tree points from the left hand side of Fig. 9 corresponds to speed of solid motion in the range of 6 – 67 μm/s, at which the averaged film thickness changes weakly, but the friction coefficient reduces significantly. The further increase of the speed of solid's motion is related with stronger increase of the averaged film thickness and insignificant change of the friction coefficient. There is obviously inverse proportion between the averaged film thickness and friction coefficient dependencies on the speed of the solid's motion.

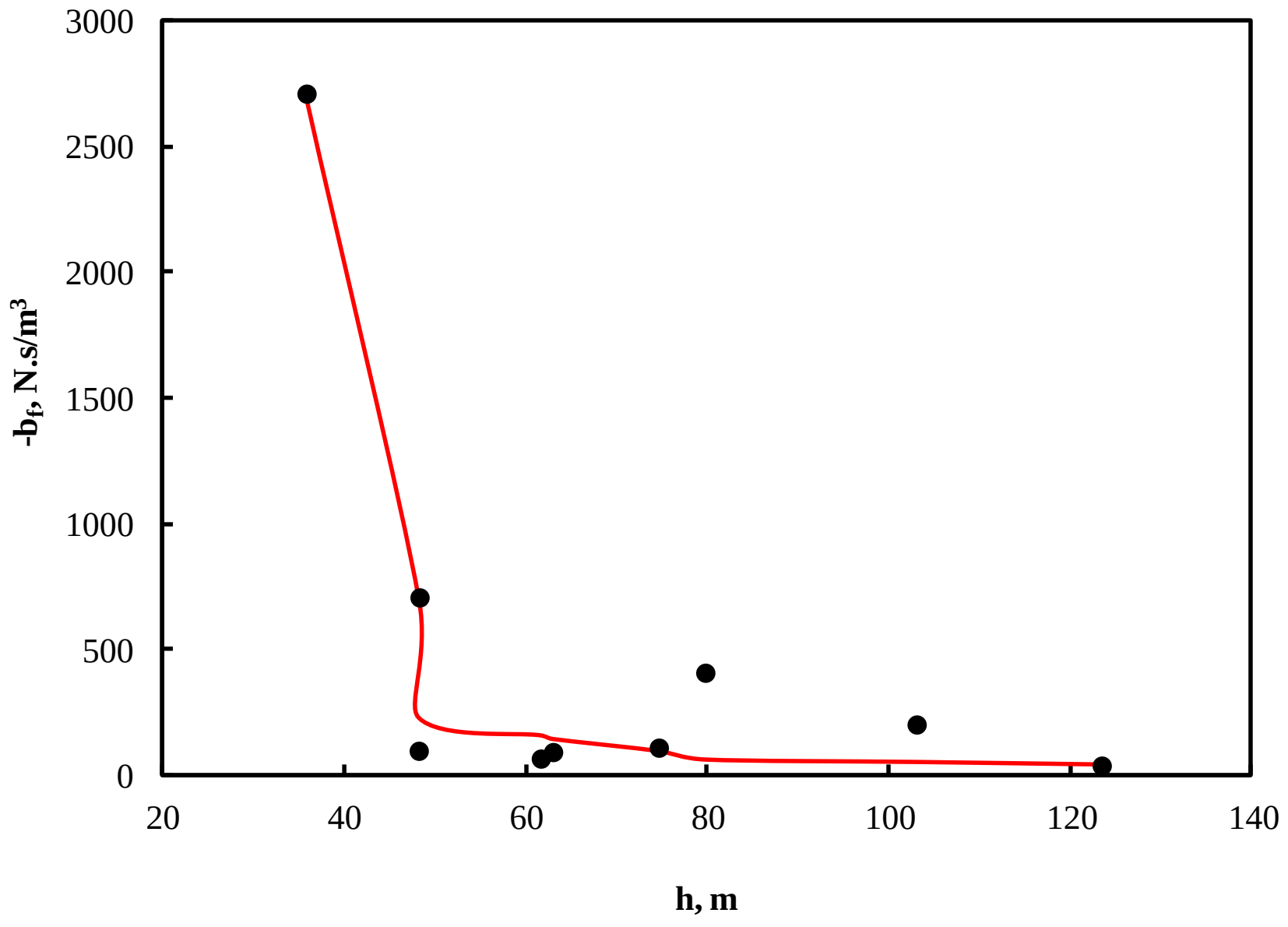

**Figure 9.** Friction coefficient per unit area $\gamma$ versus averaged film thickness: dots, obtained by experimental data and Eqs. (9) and (11); line – obtained by Eq. (10) and the relation β=1/65*U*

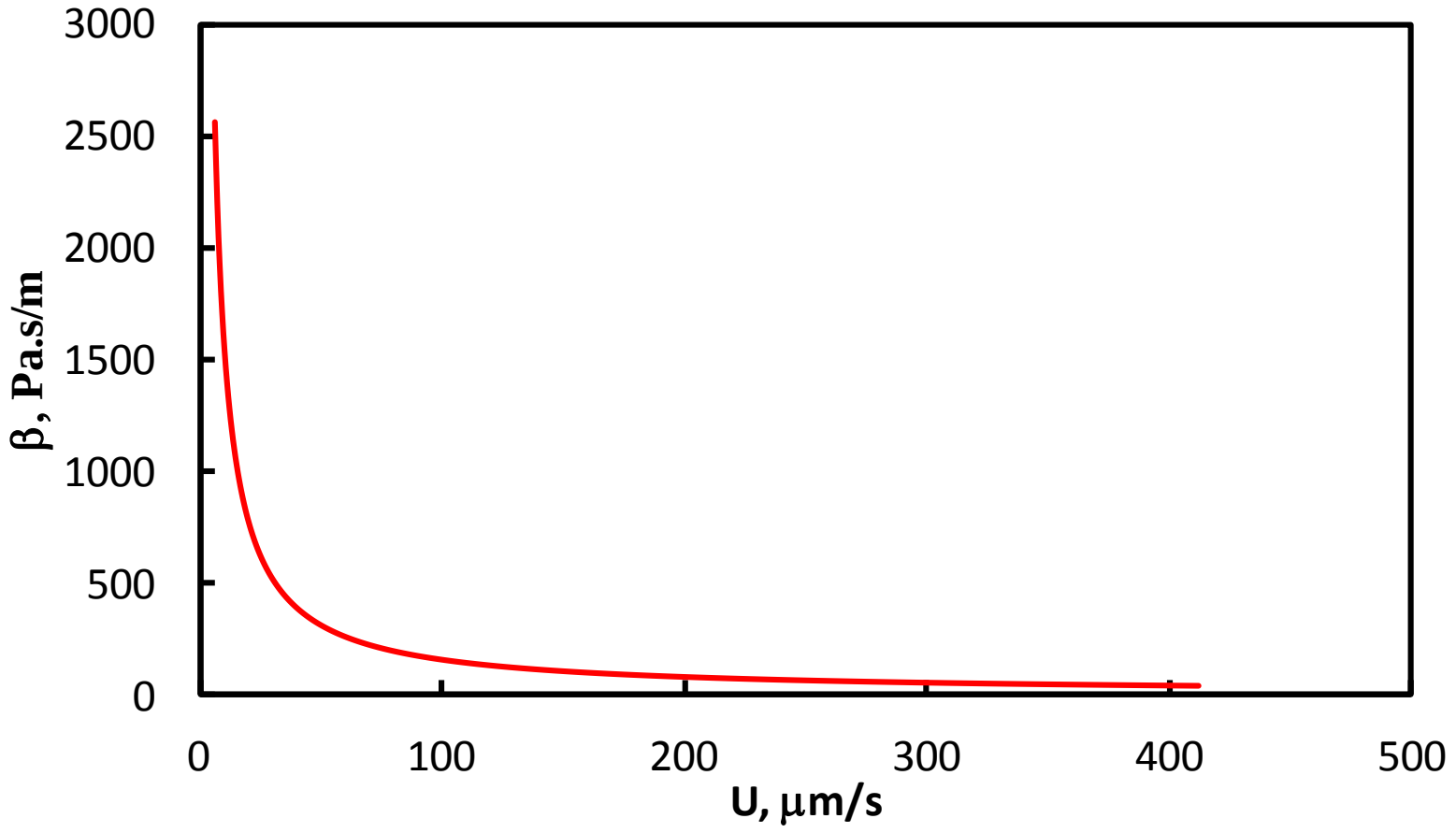

**Figure 10.** Empirical dependence of the friction slip coefficient on the speed of the solid motion, obtained via Eq. (13).

Figure 10 presents the empirical dependence of the friction slip coefficient on the speed of the solid's motion. The friction slip coefficient $\beta$ is inversely proportional to the slip between the solid and the liquid. One can see that the slip between the solid and the liquid is small at low speeds of motion. According to Fig. 10 the slip increases significantly with enhancement of the speed of the solid's motion until reaching very high values, at which almost does not depend of the speed of the solid motion anymore.

The present paper is a first step in our investigations on the tribology of viscoelastic bodies rubbing on solid surfaces. One can conclude the following: i) A regime of mixed friction between dry and lubricated frictions exists in the range of 6 – 170 $\mu$m/s, beyond which the rubbing between the bubble and solid becomes completely lubricated and passes through maximum; ii) The friction coefficient has high values at very small speeds of solid's motion and reduces and reduces substantially with the increase of the speed of the solid motion until reaching small values, which change insignificantly with the further increase of the speed of the solid; iii) It was derived empirical dependence between the friction slip coefficient and speed of the solid indicating small slip at low speeds and high slip at large speeds at which the slip becomes practically constant. The present study appears to be first one in literature devoted to the rheology of the rubbing between bubble and solid surface. Our study can be continued with further experiments with larger speeds of motion of the solid. We have to point out that 3D film thickness profiles of the wetting film at high speeds are difficult to determine as far as couple of orders of interference are usually combined on one interference pattern. Yet, such a determination is possible although being quite difficult.